# Singular polarimetry: Evolution of polarization singularities in electromagnetic waves propagating in a weakly anisotropic medium

**Konstantin Yu. Bliokh,[1,2,3]\* Avi Niv[1], Vladimir Kleiner[1], Erez Hasman[1]**

[1]*Optical Engineering Laboratory, Faculty of Mechanical Engineering,*
*Technion–Israel Institute of Technology, Haifa 32000, Israel*
[2]*Institute of Radio Astronomy, 4 Krasnoznamyonnaya St., Kharkov 61002, Ukraine*
[3]*Amcrys Ltd., 60 Lenin Ave., Kharkov, 61001, Ukraine*
*\*Corresponding author: k.bliokh@gmail.com*

**Abstract:** We describe the evolution of a paraxial electromagnetic wave characterizing by a non-uniform polarization distribution with singularities and propagating in a weakly anisotropic medium. Our approach is based on the Stokes vector evolution equation applied to a non-uniform initial polarization field. In the case of a homogeneous medium, this equation is integrated analytically. This yields a 3-dimensional distribution of the polarization parameters containing singularities, i.e. C-lines of circular polarization and L-surfaces of linear polarization. The general theory is applied to specific examples of the unfolding of a vectorial vortex in birefringent and dichroic media.

## 1. Introduction

Singular optics is an essential part of modern optics, which contributes to practically all fundamental wave phenomena [1–4]. Scalar wave fields are characterized by phase singularities (i.e., zeros of the intensity where the phase is undetermined), such as optical vortices which have found a number of applications in classical and quantum optics [3,4]. Vector fields, e.g. electromagnetic or elastic waves, have more degrees of freedom and are also characterized by *polarization singularities* [5–9]. The generic types of the polarization singularities of transverse electromagnetic waves in 3D space are C-lines and L-surfaces, where the polarizations are, respectively, circular and linear. In the two cases, the polarization ellipse degenerates either to a circle (the eccentricity vanishes and the orientation is undetermined) or to a line segment (the helicity vanishes and sign of polarization is undetermined).

The wave field singularities may form a rich variety of structures in rather simple systems. Even interference of only three plane scalar waves results in a lattice of optical vortices [10]. Clearly, propagation in inhomogeneous or/and anisotropic media significantly modifies the wave interference patterns and, hence, gives rise to a quite tangled singularities structures. Therefore, the wave field singularities in complex media are frequently studied within the *statistical* approach [7,9,11]. At the same time, for various applications it is very important to describe behavior of the specific singularities explicitly, i.e. in a *deterministic* way. There is a number of laboratory methods for generating and manipulating phase [3,4] and polarization [12,13] singularities in electromagnetic fields. Therefore, one of the currently important problems is to know how singularities evolve as the wave propagates through a medium.

Propagation of uniformly-polarized paraxial beams with phase vortices in inhomogeneous or anisotropic media have been studied recently [14–16]. While such beams represent independent localized modes of a smoothly inhomogeneous isotropic medium [14], phase vortices become drastically unstable in an anisotropic medium [15,16]. Even in the simplest uniaxial homogeneous medium, the phase vortex disappears, giving way to an essentially space-variant polarization pattern with a variety of polarization singularities [15,16]. The following features are characteristic for this system: (i) an initial phase singularity, (ii) a uniform initial polarization, and (iii) a double refraction in the medium, which destroys the phase singularity and transforms in to a set of polarization singularities.

In the present paper we aim to investigate the dynamical behavior of the polarization singularities in a paraxial wave field propagating in a weakly anisotropic and, possibly, inhomogeneous medium. However, in contrast to [15,16], the problem is considered under opposite conditions: (i) a space-variant polarization pattern with polarization singularities in the incident field and (ii) absence of the phase singularities therein; (iii) we argue that double refraction is negligible in the system, while the variations of the normal modes parameters (phases and amplitudes) along the propagation direction lead to an effective dynamics of the polarization distribution and singularities. Our choice of the initial conditions is justified by two reasons. First, it is the presence of an effective technique using subwavelength gratings for generating arbitrary space-variant polarization patterns of the field [12,13]. Second, as it follows from [15,16], the phase singularities become unstable in anisotropic media, while the polarization ones experience a continuous evolution.

By applying a dynamical approach, well-established in standard polarimetry [17,18], to a space-variant polarization pattern, we develop a powerful method for studying 3D complex polarization distributions. Assuming paraxial approximation and weak anisotropy, our approach reduces a challenging wave problem to the solution of effectively ordinary differential equation for the Stokes vector evolution along the wave propagation direction [18–26]. The equation is integrated in a homogeneous medium analytically, and, despite its simple form, it reveals an intricate evolution of the polarization singularities when the wave propagates through the medium. Thus, our method brings together polarimetry and singular optics, thereby giving rise to *singular polarimetry*. It may have promising applications – space-variant polarization patterns with singularities can be more informative and sensitive with respect to the medium properties.

## 2. General theory

### 2.1 Statement of the problem

We will examine propagation of a paraxial monochromatic electromagnetic wave through a weakly anisotropic and, possibly, inhomogeneous (stratified) medium. We assume that the wave propagates along the $z$ axis, whereas the polarization ellipse lies nearly in the $(x,y)$ plane, so that one can apply Mueller or Jones calculus to the $z$-dependent evolution of polarization [17–19]. Under this assumption, the incident field is treated as a collection of parallel rays that have essentially independent phase and amplitude evolution. Mathematically, this means that we deal with a Cauchy problem with an initial distribution of the field in the $(x,y)$ plane at $z=0$ and some dynamical equation describing the evolution of the field along the $z$ axis.

Let the polarization of the wave field at a point $\mathbf{r}$ be described through the three-component normalized Stokes vector, $\mathbf{s} = \mathbf{s}(\mathbf{r})$, $\mathbf{s}^2 = 1$, representing the polarization state on the Poincaré sphere. Then, the Cauchy problem is given by the initial Stokes vector distribution,

$$\mathbf{s}(x,y,0) = \mathbf{s}_0(x,y), \tag{1}$$

and a dynamical equation

$$\frac{\partial \mathbf{s}}{\partial z} = \hat{\mathrm{m}}\, \mathbf{s}, \tag{2}$$

where $\hat{\mathrm{m}} = \hat{\mathrm{m}}(\mathbf{s}, z)$ is a matrix operator which relates the Stokes-vectors values in two neighbor points. Although equations similar to Eq. (2) are well-established in classical polarimetry [18–26], they usually assume the uniform polarization distribution in the transverse plane, $\mathbf{s}_0(x, y) = \text{const}$, making the polarization evolution effectively *one-dimensional*, $\mathbf{s} = \mathbf{s}(z)$. In contrast, a non-uniform initial distribution (1) in our problem, $\mathbf{s}_0(x, y) \ne \text{const}$, makes the polarization distribution essentially *three-dimensional*, $\mathbf{s} = \mathbf{s}(x, y, z)$. Despite this, in order to find the complete distribution of the polarization in 3D space, $\mathbf{s}(\mathbf{r})$, one need to integrate an effectively *ordinary* differential equation (2) with initial conditions (1) for each point $(x, y)$.

The formalism of 3-component Stokes vector provides a natural representation for the polarization singularities [8,16]. Indeed, the north and south poles of the Poincaré sphere correspond to the right- and left-hand circularly polarized waves, while the equator represents linear polarizations with different orientations. Then, C- and L-type polarization singularities are determined, respectively, by the conditions

$$s_1 = 0, \; s_2 = 0, \tag{3}$$

and

$$s_3 = 0. \tag{4}$$

From Eqs. (3) and (4) it is clear that in the generic case C- and L-singularities are, respectively, lines and surfaces in 3D space: the dimension of the singularity is the dimension of the space minus the number of constraints. Alternatively, one may refer to C-points and L-lines in the $(x, y)$ plane (2D space) and their evolution along the $z$ axis. Note also, that polarization singularities are essentially determined by the third component of the Stokes vector − conditions (3) are equivalent to $|s_3| = 1$, or $s_3 = \chi$, where $\chi = \pm 1$ is the wave helicity which indicates the sign of polarization in the C-point. Having a solution of the problem Eqs. (1) and (2), $\mathbf{s} = \mathbf{s}(\mathbf{r})$, one immediately gets the space distribution of all polarization singularities from Eqs. (3) and (4). Note also that Eq. (1) implies that there are no phase singularities in the initial field, i.e. the intensity of the wave does not vanish: $I_0(x, y) = I(x, y, 0) \ne 0$ (otherwise, the Stokes vector $\mathbf{s}$ would be undefined in nodal points). As we will see, the dynamical equation (2) ensures that the nodal points cannot appear at $z \ne 0$ as well: $I(x, y, z) \ne 0$.

Our approach of $z$-dependent evolution, Eqs. (1) and (2), is justified assuming that *the refraction and diffraction processes are negligible*. Let the wave field be characterized by two scales: the wavelength $\lambda$ and a typical scale of its transverse distribution in the $(x, y)$ plane, $w \gg \lambda$. At the same time, the medium anisotropy is characterized by a typical difference between the dielectric constants corresponding to the normal modes, $\nu \ll 1$. Then, diffraction and refraction effects are negligible if: (i) the propagation distance is much smaller than the typical diffraction distance (the Rayleigh range), $z \ll z_R = w^2/\lambda$ and (ii) the propagation distance is much smaller than the distance at which the double refraction of the anisotropic medium causes transverse shifts comparable with $w$, i.e. $z \ll z_D \equiv w/\nu$. Note that the characteristic distance of the polarization evolution due to Eq. (2), $z_P = \lambda/\nu$, is much smaller than $z_D$ and can be small as compared to $z_R$. Thus, our approach is effective within the range of distances

$$z_P \leq z \ll z_R, z_D. \tag{5}$$

For instance, for a visible laser beam with $\lambda \simeq 0.6\,\mu\text{m}$ and width $w \simeq 1\,\text{mm}$ propagating through Quartz (where anisotropy is $\nu \simeq 0.03$), we have $z_R \sim 1.5\,\text{m}$, $z_D \sim 30\,\text{mm}$, and $z_P \simeq 0.02\,\text{mm}$. This gives the propagation range $0.02\,\text{mm} \leq z \ll 30\,\text{mm}$.

*2.2 Equation for the Stokes vector evolution*

To derive the evolution equation (2) for the 3-component Stokes vector $\mathbf{s}$, let us start with the 4-component Stokes vector, $\vec{S}$. Hereafter, 4-component vectors are indicated by arrows, and the last three components of a 4-vector form usual 3-component vector, so that $\vec{S} = (S_0, S_1, S_2, S_3) \equiv (S_0, \mathbf{S})$. In the most general case of a linear anisotropic medium the Stokes vector $\vec{S}$ obeys the following evolution equation [18–26]:

$$\frac{\partial \vec{S}}{\partial z} = \hat{M}\vec{S}, \tag{6}$$

where $\hat{M}$ is the differential Mueller matrix (a $4\times 4$ real matrix) which summarizes optical properties of the medium. These are given by $2\times 2$ complex dielectric tensor:

$$\hat{\varepsilon} = \varepsilon_0 \hat{I}_2 + \hat{\nu} \equiv \begin{pmatrix} \varepsilon_0 + \nu_{xx} & \nu_{xy} \\ \nu_{yx} & \varepsilon_0 + \nu_{yy} \end{pmatrix}. \tag{7}$$

Here $\varepsilon_0 \hat{I}_2$ is the main, isotropic part proportional to the unit matrix $\hat{I}_2 = \text{diag}(1,1)$, $\hat{\nu}$ is a small anisotropic part (which effectively represents the differential Jones matrix), and we assume $\text{Im}\,\varepsilon_0 = 0$ (small dissipation is ascribed to the anisotropic term). The differential Mueller matrix can be represented as [19,24,26]

$$\hat{M} = \begin{pmatrix} \text{Im}\,G_0 & \text{Im}\,G_1 & \text{Im}\,G_2 & \text{Im}\,G_3 \\ \text{Im}\,G_1 & \text{Im}\,G_0 & -\text{Re}\,G_3 & \text{Re}\,G_2 \\ \text{Im}\,G_2 & \text{Re}\,G_3 & \text{Im}\,G_0 & -\text{Re}\,G_1 \\ \text{Im}\,G_3 & -\text{Re}\,G_2 & \text{Re}\,G_1 & \text{Im}\,G_0 \end{pmatrix}. \tag{8}$$

Here the complex 4-component vector $\vec{G} = (G_0, \mathbf{G})$ is expressed via components of the dielectric tensor (7) as

$$\vec{G} = -\frac{k_0}{2\sqrt{\varepsilon_0}} \left[ (\nu_{xx} + \nu_{yy}), (\nu_{xx} - \nu_{yy}), (\nu_{xy} + \nu_{yx}), i(\nu_{xy} - \nu_{yx}) \right]^T, \tag{9}$$

where $k_0$ is the wave number in vacuum, and components of quantities (7)–(9) can be $z$-dependent. Vector $\vec{G}$ gives decomposition of the anisotropy tensor $\hat{\nu}$, Eq. (6), with respect to the basis of Pauli matrices, and establishes close relations between polarization optics (Mueller and Jones calculus) and relativistic problems with the Lorentz-group symmetry [20,24,26–31]. In particular, Eq. (6) for the Stokes vector evolution is similar to the Bargman–Michel–Telegdi equation for relativistic spin precession [26,32].

The matrix $\hat{M}$ can be decomposed into three parts responsible for different optical properties of the medium [18–26]:

$$\hat{M} = \mathrm{Im}\,G_0 \hat{I}_4 + \begin{pmatrix} 0 & \mathrm{Im}\,G_1 & \mathrm{Im}\,G_2 & \mathrm{Im}\,G_3 \\ \mathrm{Im}\,G_1 & 0 & 0 & 0 \\ \mathrm{Im}\,G_2 & 0 & 0 & 0 \\ \mathrm{Im}\,G_3 & 0 & 0 & 0 \end{pmatrix} + \begin{pmatrix} 0 & 0 & 0 & 0 \\ 0 & 0 & -\mathrm{Re}\,G_3 & \mathrm{Re}\,G_2 \\ 0 & \mathrm{Re}\,G_3 & 0 & -\mathrm{Re}\,G_1 \\ 0 & -\mathrm{Re}\,G_2 & \mathrm{Re}\,G_1 & 0 \end{pmatrix}, \qquad (8')$$

where $\hat{I}_4 = \mathrm{diag}(1,1,1,1)$ is the unit matrix. The first, diagonal part, proportional to $\mathrm{Im}\,G_0$, describes the common *attenuation* of the field intensity. The second, symmetric part, related to components of $\mathrm{Im}\,\mathbf{G}$, describes the phenomenon of *dichroism*, i.e. the selective attenuation of different field components. Finally, the third, antisymmetric part of $\hat{M}$, related to components of $\mathrm{Re}\,\mathbf{G}$, is responsible for the medium *birefringence*. The component $\mathrm{Re}\,G_0$ does not contribute to the matrix (8), since it causes merely an additional total phase of the wave field, which does not affect the polarization state and is lost in the Stokes vector representation.

Evolution of the normalized 3-component Stokes vector can be derived immediately from Eqs. (6) and (8) by differentiating the definition $\mathbf{s} = \mathbf{S}/S_0$. As a result, we find that the 3-component Stokes vector obeys the following equation [26]:

$$\frac{\partial \mathbf{s}}{\partial z} = (\mathbf{\Omega} + \mathbf{s} \times \mathbf{\Sigma}) \times \mathbf{s}, \qquad (10)$$

where we denoted $\mathrm{Re}\,\mathbf{G} \equiv \mathbf{\Omega}$ and $\mathrm{Im}\,\mathbf{G} \equiv \mathbf{\Sigma}$. This equation represents the basic evolution equation (2) in the general case. [In terms of Eq. (2), the matrix $\hat{m}$ is given by $\mathrm{m}_{ij} = -e_{ijk}\Omega_k - e_{ijk}e_{klm}s_l\Sigma_m$, where indices take values $1,2,3$ and $e_{ijk}$ is the unit antisymmetric tensor.] Thus, all the evolution on the Poincaré sphere can be described by the precession equation (10) which includes two real vectors $\mathbf{\Omega}$ and $\mathbf{\Sigma}$ responsible for the birefringent and dichroic effects, respectively. Equation (10) conserves the absolute value of the normalized Stokes vector under the evolution: $\partial \mathbf{s}^2/\partial z = 0$. The common attenuation, $\mathrm{Im}\,G_0$, naturally, does not affect the normalized Stokes vector and is absent in Eq. (10). Note also that Eq. (10) resembles the Landau–Lifshitz equation describing the nonlinear spin precession in ferromagnets [33], but, in contrast to the latter, Eq. (10) contains two *different* effective fields $\mathbf{\Omega}$ and $\mathbf{\Sigma}$. In inhomogeneous medium $\mathbf{\Omega} = \mathbf{\Omega}(z)$ and $\mathbf{\Sigma} = \mathbf{\Sigma}(z)$.

*2.3 Solutions in a homogeneous birefringent medium*

In a homogeneous non-dissipative birefringent medium, Eq. (10) takes simple form of the classical precession equation [22,23,25]:

$$\frac{\partial \mathbf{s}}{\partial z} = \mathbf{\Omega} \times \mathbf{s}. \qquad (11)$$

According to Eq. (11), as the wave propagates along the $z$ axis, the Stokes vector $\mathbf{s}$ precesses with a constant spatial frequency $\Omega$ about the fixed direction $\boldsymbol{\omega} = \mathbf{\Omega}/\Omega$. In terms of the medium properties, direction $\boldsymbol{\omega}$ and absolute value $\Omega$ characterize, respectively, the type and the strength of the medium birefringence. In so doing, two "stationary" solutions $\mathbf{s}^{\pm} = \pm\boldsymbol{\omega}$ on the Poincaré sphere correspond to mutually-orthogonal eigenmodes of the medium. In particular, $\boldsymbol{\omega} = (0,0,\pm 1)$ and $\boldsymbol{\omega} = (\omega_1,\omega_2,0)$ correspond, respectively, to the cases of circularly- and linearly-birefringent medium. Equation (11) with initial condition (1) can be easily integrated: $\mathbf{s}(x,y,z) = \hat{R}_\omega(\Omega z)\mathbf{s}_0(x,y)$, where $\hat{R}_\omega(\Omega z)$ is the operator of rotation about $\boldsymbol{\omega}$ on the angle $\Omega z$. Using the Rodrigues rotation formula [34], we arrive at

$$\mathbf{s} = \mathbf{s}_0 \cos(\Omega z) + (\boldsymbol{\omega} \times \mathbf{s}_0)\sin(\Omega z) + (\boldsymbol{\omega}\mathbf{s}_0)\boldsymbol{\omega}[1 - \cos(\Omega z)]. \tag{12}$$

Together with Eq. (1), this equation gives solution for the Stokes-vector distribution in space.

In a circularly birefringent medium, distribution of polarization singularities in $(x,y)$ plane does not vary when the wave propagates along the $z$ axis. Indeed, the helicity of the wave, given by the third component of the Stokes vector, is invariant of Eqs. (11) and (12), $s_3 = \text{const}$, when $\boldsymbol{\omega} = (0,0,\pm 1)$. Thus, C-lines ($s_3 = \pm 1$) and L-surfaces ($s_3 = 0$) are parallel to the $z$ axis in this case and are trivially determined by the initial distribution (1) with Eqs. (3) and (4):

$$s_{01} = 0, \quad s_{02} = 0, \tag{13}$$

and

$$s_{03} = 0. \tag{14}$$

On the contrary, polarization singularities evolve in a linearly-birefringent medium, cf. [15,16]. As it is clear from Eq. (12), this evolution is periodic in $z$ with the period $2\pi/\Omega$. In fact, L-lines and C-points in $(x,y)$ plane come back to the initial locations after $\pi/\Omega$ period, corresponding to the half-wavelength plate. In so doing, C-points only change their signs after $\pi/\Omega$ period. One can also note that under propagation at $\pi/2\Omega$ distance (corresponding to the quarter-wavelength plate) C-points give their place to points of L-lines, while some points of L-lines give place to C-points. To determine the whole 3D structure of polarization singularities note that vector $\boldsymbol{\omega} = (\omega_1, \omega_2, 0)$ can be reduced to $\boldsymbol{\omega} = (1,0,0)$ by a fixed rotation of coordinate axes in the $(x,y)$ plane, which brings the anisotropy tensor $\hat{\nu}$, Eq. (7), to the principal axes. Then, substituting solution (12) with $\boldsymbol{\omega} = (1,0,0)$ into Eqs. (3) and (4), we obtain equations determining C-lines and L-surfaces:

$$s_{01} = 0, \quad \tan(\Omega z) = \frac{s_{02}}{s_{03}}, \tag{15}$$

and

$$\tan(\Omega z) = -\frac{s_{03}}{s_{02}}. \tag{16}$$

In contrast to Eqs. (13) and (14), these rather simple equations reveal non-trivial $z$-dependent dynamics of polarization singularities (see examples in Section 3.2).

*2.4 Solutions in a homogeneous dichroic medium*

In a homogeneous dichroic medium, with selective attenuation of modes but without a phase difference between them (i.e., $\boldsymbol{\Omega} = 0$), Eq. (10) takes the form

$$\frac{\partial \mathbf{s}}{\partial z} = (\mathbf{s} \times \boldsymbol{\Sigma}) \times \mathbf{s}, \tag{17}$$

Similarly to Eq. (11), this equation has two "stationary" solutions $\mathbf{s}^\pm = \pm\boldsymbol{\sigma}$ (where $\boldsymbol{\sigma} = \boldsymbol{\Sigma}/\Sigma$), which determine eigenmodes of the medium. However, in contrast to the birefringent-medium case, solution $\mathbf{s}^-$ is "unstable". As we will see, solutions of Eq. (17) move on the Poincaré sphere away from $\mathbf{s}^-$ towards $\mathbf{s}^+$. Thus, the dichroic medium is a polarizer, in which only one mode (given by $\mathbf{s}^+$) survives at long enough propagation distances. Equation (17) can be integrated analytically at $\Sigma = \text{const}$, which yields the solution (see Appendix):

$$\mathbf{s} = \frac{2}{(1+A_0)e^{\Sigma z}+(1-A_0)e^{-\Sigma z}}\mathbf{s}_0 + \frac{(1+A_0)e^{\Sigma z}-2A_0-(1-A_0)e^{-\Sigma z}}{(1+A_0)e^{\Sigma z}+(1-A_0)e^{-\Sigma z}}\boldsymbol{\sigma}, \tag{18}$$

where $A_0 = \mathbf{s}_0\boldsymbol{\sigma}$. As seen from Eq. (18), all solutions with $\mathbf{s}_0 \neq \mathbf{s}^-$ move in the plane given by vectors $\mathbf{s}_0$ and $\boldsymbol{\sigma}$, and tend exponentially to $\mathbf{s}^+$ point. Supplied with the initial distribution (1), Eq. (18) gives the Stokes-vector distribution in 3D space.

In contrast to the birefringent-medium case, polarization singularities vary along $z$ both in circularly and linearly dichroic medium. More precisely, in a circularly-dichroic medium, $\boldsymbol{\sigma} = (0,0,\pm 1)$, C-points in the $(x,y)$ plane do not evolve with $z$ since circular polarizations correspond to "stationary" solutions $\mathbf{s}^\pm$ in this case. Thus C-lines in 3D space are parallel to the $z$ axis and determined analogously to Eq. (13):

$$s_{01} = 0, \; s_{02} = 0. \tag{19}$$

At the same time, L-lines in the $(x,y)$ plane evolve with $z$, and L-surfaces in 3D space are determined by Eq. (4) with (18):

$$\frac{(1+A_0)e^{\Sigma z}-(1-A_0)e^{-\Sigma z}}{(1+A_0)e^{\Sigma z}+(1-A_0)e^{-\Sigma z}} = 0, \tag{20}$$

where $A_0 = s_{03}\sigma_3$. Equation (20) can be resolved with respect to $z$, which yields

$$\tanh(\Sigma z) = -s_{03}\sigma_3. \tag{21}$$

Conversely, in a linearly-dichroic medium, where $\boldsymbol{\sigma} = (1,0,0)$ in the principal-axes coordinate frame, L-surfaces are parallel to the $z$ axis and are given by Eq. (14):

$$s_{03} = 0. \tag{22}$$

At the same time, C-points in the $(x,y)$ plane evolve with $z$, and 3D C-lines are determined by Eq. (3) with (18), i.e.

$$\frac{(1+A_0)e^{\Sigma z}-(1-A_0)e^{-\Sigma z}}{(1+A_0)e^{\Sigma z}+(1-A_0)e^{-\Sigma z}} = 0, \; s_{02} = 0, \tag{23}$$

where $A_0 = s_{01}$. Resolving of this equation yields:

$$\tanh(\Sigma z) = -s_{01}, \; s_{02} = 0. \tag{24}$$

Thus, unlike birefringent medium, evolution of the Stokes vector and polarization singularities in dichroic medium is monotonic rather than periodic, see examples in Section 3.3. It is described by hyperbolic functions, which appear naturally in the Lorentz-group representation of polarization optics [30,31].

## 3. Application: Evolution of a vectorial vortex

*3.1 Initial polarization distribution*

As a characteristic example of initial space-variant polarization pattern, Eq. (1), we consider a *vectorial vortex*, which possesses a singularity in the polarization distribution [12,13]. The Stokes-vector distribution (1) of a vectorial vortex at the origin can be given as

$$s_{01} = \sqrt{1-f^2(\rho)}\cos[m(\varphi-\delta)],$$
$$s_{02} = \sqrt{1-f^2(\rho)}\sin[m(\varphi-\delta)], \qquad (25)$$
$$s_{03} = f(\rho).$$

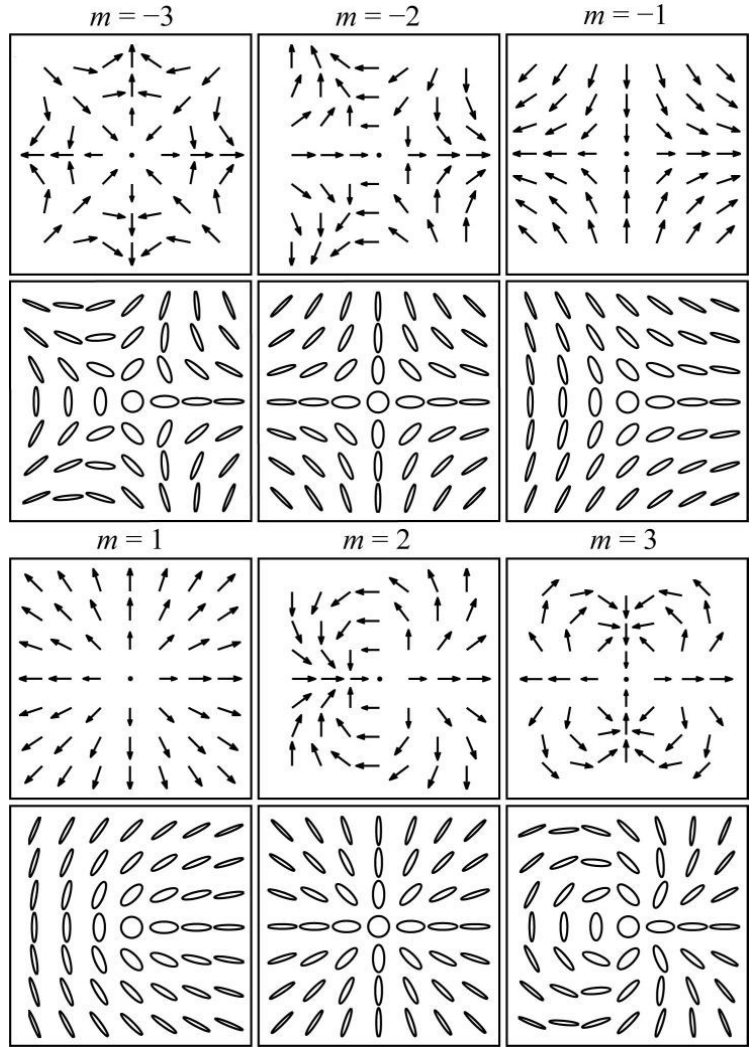

Fig. 1. Distributions of Stokes vectors (upper panel) and of polarization ellipses (lower panel) for vectorial vortices Eqs. (25) with a C-point in the center, Eq. (26), at different values of azimuthal index $m$. Hereafter, $\delta = 0$ and directions of Stokes vectors are naturally depicted in the real space with the $s_1$, $s_2$, and $s_3$ components pointing along the $x$, $y$, and $z$ axis, respectively.

Here $(\rho,\varphi)$ are the polar coordinates in the $(x,y)$ plane, $f(\rho)$ is a radial distribution function, $|f(\rho)| \leq 1$, $m = \pm 1, \pm 2, ...$ is the integer number (the azimuthal index of the polarization distribution), and $\delta$ indicates a fixed angle between the distribution and $x$ axis. In the above distribution, the Stokes vector experiences $m$ complete rotations along a loop path enclosing the vortex center. Therefore, the distribution possesses the $|m-1|$-fold rotational symmetry (one turn is effectively compensated by a $2\pi$ rotation of local radial vector), Fig. 1. At the same time, the corresponding polarization pattern (i.e. the distribution of polarization ellipses in the $(x,y)$ plane) reveals $|m-2|$-fold symmetry, Fig. 1. This is because a complete turn of the Stokes vector corresponds to a half-turn of the polarization ellipse; as a result, the symmetry of the polarization distribution is characterized by the order of $|m\pi - 2\pi| \mod \pi$. Note that the cases $m = 1$ and $m = 2$ are peculiar: the vectorial vortex represents an azimuthally-symmetric Stokes-vector and polarization distributions, respectively.

Points $\rho = \rho_C$ and $\rho = \rho_L$, such that $f(\rho_C) = \chi = \pm 1$ and $f(\rho_L) = 0$ correspond to C- and L-type singularities in the initial distribution (25). We will concentrate on a simple case of a single C-point with the right-hand circular polarization at the origin, so that $f(0) = \chi$ and $0 < |f(\rho)| < 1$ at $\rho > 0$. This case can be modeled using the function

$$f = \frac{\chi}{\sqrt{1 + (\rho/\rho^*)^2}}, \tag{26}$$

where $\rho^*$ characterizes the radial scale of the distribution. The vectorial vortex (25) and (26) with the right-hand polarization in the center, $\chi = 1$, is characterized by the optical vortex (phase singularity) in the left-hand polarized component of the field, and vice versa. Below, we will examine the behavior of polarization singularities in homogeneous anisotropic media considered in Sections 2.3 and 2.4, assuming the initial polarization distribution of Eqs. (25) and (26).

*3.2 Homogeneous linearly-birefringent medium*

Since the behavior of polarization singularities in a circularly-birefringent medium is trivial, Eqs. (13) and (14), let us consider the case of linearly-birefringent medium. Substituting the initial Stokes-vector distribution, Eqs. (25), into Eqs. (15) and (16) we obtain the equations describing C-lines and L-surfaces in space. For C-lines, this yields

$$\sqrt{1 - f(\rho)^2} \cos[m(\varphi - \delta)] = 0, \quad \tan(\Omega z) = \frac{\sqrt{1 - f^2(\rho)} \sin[m(\varphi - \delta)]}{f(\rho)}, \tag{27}$$

whereas L-surfaces are described by equation

$$\tan(\Omega z) = -\frac{f(\rho)}{\sqrt{1 - f^2(\rho)} \sin[m(\varphi - \delta)]}. \tag{28}$$

Assuming the initial distribution with radial function (26), we find that solutions of Eqs. (27) represent curves lying in the azimuthal planes $\varphi = \text{const}$ and given by equations

$$\varphi_n = \delta + \frac{\pi(2n+1)}{2m}, \quad \tan(\Omega z) = \pm \frac{\sqrt{1-f^2(\rho)}}{f(\rho)} = \pm \chi \frac{\rho}{\rho^*}. \tag{29}$$

Here $n = 0, 1, ..., 2m-1$, signs "$\pm$" in the second equation correspond to even and odd $n$, respectively. Note that $\tan(\Omega z)$ is either positive or negative at each value of $z$, and, hence, only solutions with either even or odd $n$ are valid each time. They alternate after a period of $\pi/2\Omega$, whereas all the structure of polarization singularities (up to sign of the polarization) has a period of $\pi/\Omega$. L-surfaces (28) separate C-lines with different helicities $\chi$ (and space areas with positive and negative $s_3$) and represent azimuthally-corrugated surfaces. Figure 2 shows an example of the polarization singularities described by Eqs. (29) and (28). The initial C-point of $m$ th order splits into $m$ branches under the evolution along $z$. This reveals an *instability* of higher-order C-points during evolution in an anisotropic medium. Only C-points with of a minimal order, $m = \pm 1$, are generic [5–9].

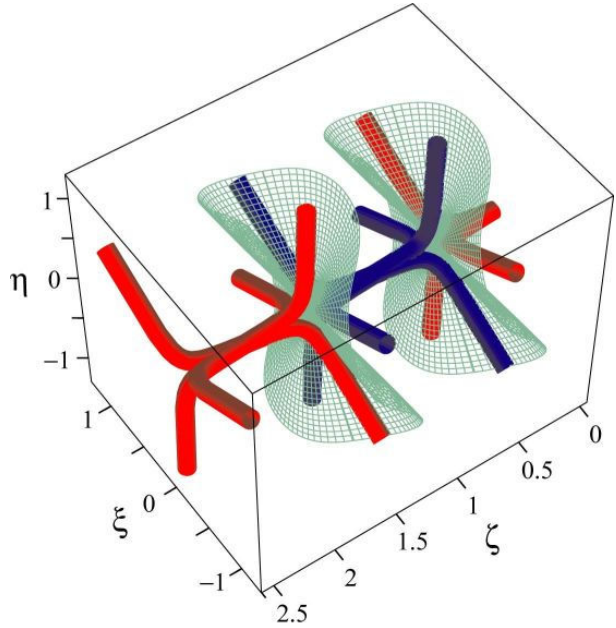

Fig. 2. C-lines and L-surfaces under propagation of the vectorial vortex, Eqs. (25) and (26), with $m = 3$ and $\chi = 1$ in a linearly-birefringent medium with $\boldsymbol{\omega} = (1,0,0)$. Dimensionless coordinates $\xi = x/\rho^*$, $\eta = y/\rho^*$, and $\zeta = \Omega z/\pi$ are used, whereas red and blue colors indicate C-lines with $\chi = 1$ and $\chi = -1$, respectively.

Similarly to [15], one can verify conservation of the total *topological charge* by tracking the polarization singularities evolution. Topological charge, $\gamma$, characterizes each C-point in plane $z = $ const and is equal to the product of its local azimuthal index, $m$, and sign of the polarization, $\chi = s_3$: $\gamma = m\chi$. The total topological charge, $\Gamma = \sum_a \gamma^{(a)}$, should be $z$-independent [7,9,15]:

$$\Gamma = \sum_a m^{(a)} \chi^{(a)} = \text{const}. \tag{30}$$

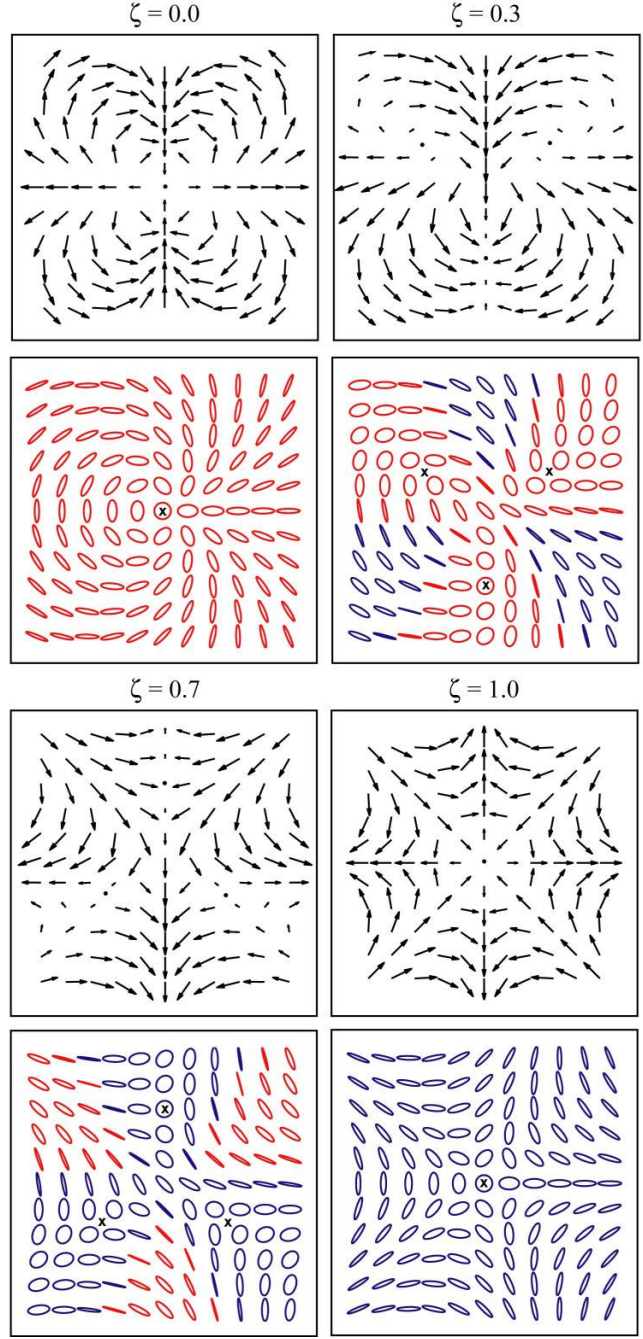

Fig. 3. Distributions of Stokes vectors, Eq. (12), (upper panel) and of polarization ellipses (lower panel) for vectorial vortex, Eqs. (25) and (26), with $m=3$ and $\chi=1$, propagating in a linearly-birefringent medium with $\boldsymbol{\omega}=(1,0,0)$, Fig. 2. Distributions are shown at different propagation distances $\zeta=\Omega z/\pi$ within half-period. Red and blue colors indicate areas with right-hand ($s_3>0$) and left-hand ($s_3<0$) polarizations. C-points are marked by dots (upper panel) and crosses (lower panel).

Here $a$ indicates different C-points and summation is taken over whole plane $z = \text{const}$. Figure 3 demonstrates polarization and Stokes-vector distributions, Eq. (12), for a vectorial vortex evolving in a linearly birefringent medium corresponding to Fig. 2. Conservation of $\Gamma = 3$ is seen – initial C-point with $m = 3$ and $\chi = 1$ is split into three C-points with $m = 1$ and $\chi = 1$; then, crossing of L-surface causes simultaneous flip of helicity and azimuthal index: three C-points with $m = -1$ and $\chi = -1$ occur; finally, they merge into single C-point with $m = -3$ and $\chi = -1$.

*3.3 Homogeneous dichroic media*

Substituting initial distribution Eqs. (25) into Eqs. (18)–(24), we get the Stokes-vector distribution and polarization singularities in homogenous dichroic medium. In a circularly-dichroic medium, C-points do not evolve with $z$, Eq. (19), whereas the behavior of L-surfaces is given by Eq. (21) with Eq. (25):

$$\tanh(\Sigma z) = -f(\rho)\sigma_3. \tag{31}$$

Thus, the surface lies in the $z < 0$ or $z > 0$ half-space when $\sigma_3 > 0$ or $\sigma_3 < 0$, respectively. Structure of the polarization singularities with initial distribution (26) in a circularly-dichroic medium is shown in Fig. 4a.

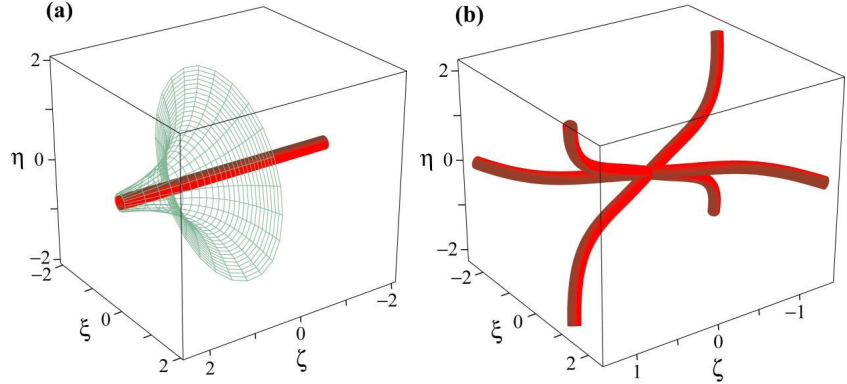

Fig. 4. C-lines and L-surfaces under propagation of the vectorial vortex, Eqs. (25) and (26), with $m = 3$ and $\chi = 1$ in a dichroic medium. Pictures (a) and (b) correspond to circularly- and linearly-dichroic media with $\boldsymbol{\sigma} = (0, 0, -1)$ and $\boldsymbol{\sigma} = (1, 0, 0)$, respectively. Dimensionless coordinates $\xi = x/\rho^*$, $\eta = y/\rho^*$, and $\zeta = \Sigma z$ are used.

In a linearly-dichroic medium, L-surfaces are trivial, Eq. (22), whereas C-lines are described by Eqs. (24) with Eq. (25):

$$\tanh(\Sigma z) = -\sqrt{1 - f^2(\rho)}\cos[m(\varphi - \delta)], \quad \sqrt{1 - f^2(\rho)}\sin[m(\varphi - \delta)] = 0. \tag{32}$$

Assuming initial distribution with the radial function (26), we find that, similarly to Eq. (29), solutions of Eqs. (32) represent curves lying in the azimuthal planes $\varphi = \text{const}$:

$$\varphi_n = \delta + \frac{\pi n}{m}, \quad \tanh(\Sigma z) = \mp\sqrt{1 - f^2(\rho)}, \tag{33}$$

Here $n = 0, 1, \ldots, 2m - 1$, and signs "$\mp$" in the second equation correspond to even and odd $n$, respectively. It is easily seen that solutions with even and odd $n$ are realized, respectively, at

$z < 0$ and $z > 0$, Fig. 4b. Thus, similarly to the case of linearly-birefringent medium, the initial C-point of the $m$ th order is split into $m$ C-points with unit azimuthal indices, Fig. 5. The total topological charge, Eq. (30), is also conserved in this process.

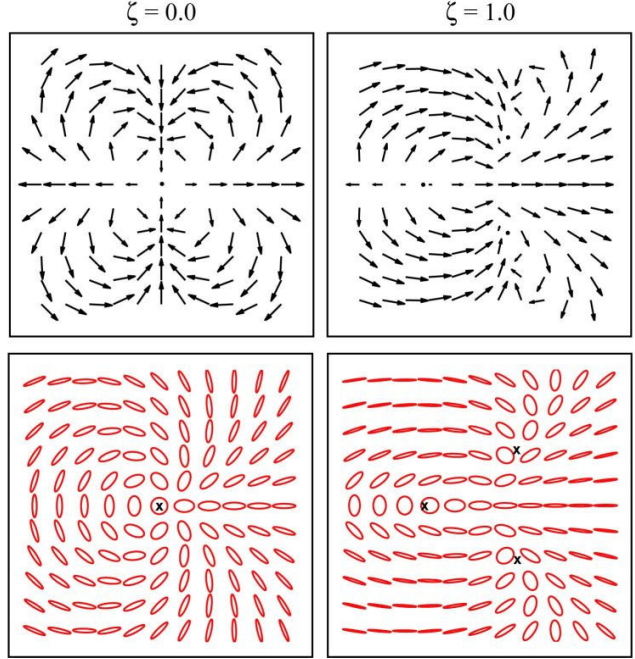

Fig. 5. Distributions of Stokes vectors, Eq. (18), (upper panel) and of polarization ellipses (lower panel) for vectorial vortex, Eqs. (25) and (26), with $m = 3$ and $\chi = 1$, propagating in a linearly-dichroic medium with $\boldsymbol{\sigma} = (1,0,0)$, Fig. 4b. Distributions are shown at different propagation distances $\zeta = \Sigma z$. C-points are marked by dots (upper panel) and crosses (lower panel).

## 4. Conclusion

To summarize, we have developed an efficient formalism describing the evolution of non-uniformly polarized waves in anisotropic media. Provided that refraction and diffraction effects are negligible, our method reduces the initial wave problem to the integration of a simple differential equation for the Stokes-vector evolution. Polarization singularities are readily found in the resulting space distribution of the Stokes vector. The evolution equation has been integrated analytically for the characteristic cases of homogeneous birefringent and dichroic media.

    We have applied the general formalism to the evolution of a polarization vortex in birefringent and dichroic media. The resulting space polarization patterns describe remarkable behavior of polarization singularities as the wave propagates in the medium. In particular, we showed the splitting of a higher-order vectorial vortex into a number of the minimal-order generic vortices and verified conservation of the topological charge under that process.

    The fine behavior of the wave-field singularities in anisotropic media can be used for needs of polarimetry, and, perhaps, will enable one to increase the sensitivity of classical polarimetric methods.

    Finally, though we considered a rectilinear propagation of the wave along the $z$ axis, our approach is also valid for the geometrical-optics wave propagation along smooth curvilinear rays in large-scale inhomogeneous media. In this case, the problem is reduced to the same

form if one involves a local ray coordinate system with basis vectors being parallel-transported along the ray [26].

*Note added.*– Recent paper [35] with related arguments came to our attention after submission of this work.

**Acknowledgements**

This work was partially supported by STCU (grant P-307) and CRDF (grant UAM2-1672-KK-06).

**Appendix: Solution of equation (17)**

To integrate Eq. (17), note that the unit vector **s** can be represented as

$$\mathbf{s} = A\boldsymbol{\sigma} - (\mathbf{B} \times \boldsymbol{\sigma}). \tag{A1}$$

Here we introduced two auxiliary quantities: scalar $A = \mathbf{s}\boldsymbol{\sigma}$ and vector $\mathbf{B} = (\mathbf{s} \times \boldsymbol{\sigma})$. From Eq. (17), it can be easily seen that they obey equations

$$\frac{\partial A}{\partial z} = \Sigma(1 - A^2), \tag{A2}$$

$$\frac{\partial \mathbf{B}}{\partial z} = -\Sigma A \mathbf{B}. \tag{A3}$$

By integrating Eq. (A2) we obtain

$$A = \frac{(1 + A_0)e^{\Sigma z} - (1 - A_0)e^{-\Sigma z}}{(1 + A_0)e^{\Sigma z} + (1 - A_0)e^{-\Sigma z}}, \tag{A4}$$

where $A_0 = A|_{z=0}$. Substituting Eq. (A4) into Eq. (A3) and performing integration, we arrive at solution for **B**:

$$\mathbf{B} = \frac{2}{(1 + A_0)e^{\Sigma z} + (1 - A_0)e^{-\Sigma z}} \mathbf{B}_0, \tag{A5}$$

where $\mathbf{B}_0 = \mathbf{B}|_{z=0}$. Substituting Eqs. (A4) and (A5) into Eq. (A1) yields solution (18).